\documentclass[reprint,12pt]{elsarticle}
\usepackage{amssymb}
\usepackage{bm}
\usepackage{ulem}
\usepackage{xcolor}
\usepackage{lineno}

\begin{document}

\begin{frontmatter}

\title{Modeling the precipitation distribution by radius and pore size during drying of an impregnated sphere}
 
\author{N.V. Peskov}
\ead{peskovnick@gmail.com}
\address{Faculty of Computational Mathematics and Cybernetics, \\Lomonosov Moscow State University,  Moscow, Russian Federation.}

\begin{abstract}
The process of preparing heterogeneous catalysts on porous supports includes a drying stage, in which the porous material, impregnated with an aqueous solution of the catalyst precursor, is dried, and the precursor is precipitated on the pore walls. The precipitate  distribution throughout the support volume strongly influences the catalyst performance and durability within industrial reactors.
This paper presents a mathematical model for simulating the precipitation distribution during the drying of a porous sphere. A method for calculating the precipitant distribution by pore size is proposed. Examples of calculations are given demonstrating the influence of some parameters on the distribution by space and by pore size.

\end{abstract}

\end{frontmatter}
%\linenumbers

\section{Introduction}

Heterogeneous catalysts are widely used in the modern chemical industry to accelerate reactions and alter process selectivity in the desired direction. Improving catalyst efficiency can be achieved, in particular, by increasing the contact area of its active elements with the reactants. This is achieved by distributing the active elements over the pore surface of a porous support with a large internal surface area. One of the most common methods for preparing such catalysts is impregnation, in which the support is first impregnated with a precursor solution, then dried and calcined  \cite{Mun15, JUB20}. In industrial reactors, such catalysts are used in the form of millimeter-sized granules of various shapes, of which spherical ones are the most convenient for modeling. 

In known models for preparing supported catalysts by impregnation, two mechanisms for depositing active elements on pore walls were considered. Chronologically, the first mechanism to be considered was one in which active elements were adsorbed from liquid onto the pore walls. In this case, the distribution of active elements in the pore space was mainly formed at the impregnation stage, and subsequent operations only corrected this distribution \cite{VM74, LA85, L01, LKG08}. This mechanism works in cases where the salt concentration in the liqud is low and the chemical interaction between the solute and solid support is strong enough. In another mechanism, the catalyst precursor in the form of a catalyst metal salt is deposited from the supersaturated aqueous solution on the pore walls during the drying step \cite{RPK1, RPK2}.

In this paper, a mathematical model of the second mechanism of placement of the catalyst precursor in the pore space of the support is presented. This model, like model \cite{RPK1, RPK2}, is based on the Whitaker model of drying porous materials \cite{SW, VT19} with additional terms describing the precipitation of salt from liquid in the pore space. However, the equation for the gas phase is excluded from the equations of model \cite{SW}, since precipitation occurs only in the presence of a liquid in the pores. In this case, dynamic equilibrium between the liquid and vapor is usually assumed, that is, the vapor density is equal to the saturated density. Under these conditions, the equation for the gas phase can be replaced by the requirement that the gas pressure in the pores be equal to the external gas pressure. This substitution has virtually no effect on the overall solution of the model.

The results of calculating the distribution of precipitant along the radius of a sphere in dependence on some model parameters are presented. To simulate precipitant distribution over pores, a capillary bundle model is used, which is used in the Whitaker model to estimate capillary pressure and fluid velocity within pores. It is assumed that precipitation settles in those pores (capillaries) that are currently drying.

\section{Model}

A homogeneous porous sphere with radius $R$ and porosity $\epsilon$ = $V_p/V$, where $V$ is the volume of the sphere and $V_p$ is the total volume of the pores (voids), is considered. The sphere was impregnated with an aqueous solution of a metal salt (some physical properties of ferrous sulfate FeSO$_4\cdot$7(H$_2$O) will be referred), which serves as a precursor for the catalyst. It is assumed that the solution does not chemically interact with the sphere material and that salt adsorption from the solution onto the pore walls can be neglected.
After impregnation, the sphere is dried in air under specified external conditions: temperature $T_{ex}$, pressure $P_{g,ex}$, and humidity $\varphi_{ex}$. The corresponding water vapor pressure in the external air is calculated using the formula $P_{v,ex} = \varphi_{ex}P^*_v(T_{ex})$, where $P_v^*(T)$ is the saturated water vapor pressure at temperature $T$. During the drying process, water evaporates from the liquid, increasing the salt concentration, and when the solubility threshold is exceeded, the salt precipitates on the pore walls.

The mathematical model presented below for drying a liquid in a porous sphere is based on the Whitaker model \cite{SW} with additional terms that take into account the precipitation of the dissolved substance from a supersaturated solution.

\subsection{Equations}

The Whitaker model describes both the drying of liquid mobile moisture in pores and the drying of immobile moisture bound to the solid skeleton of the material. In this study, only the drying of mobile moisture -- the solution from which the salt precipitates -- is of interest. Therefore, the Whitaker system of equations is considered only in the presence of liquid in the pores. In this case, the solvability of the system does not require invoking the sorption isotherm of the material. It is sufficient to assume dynamic equilibrium between the liquid and water vapor in the pores. That is, assume that the vapor density $\rho_v$ is always equal to the saturated vapor density $\rho_v^*$.

The continuity equation for a liquid in the Whitaker system for an aqueous solution can be split into two equations: one for the aqueous portion of the liquid and one for the dissolved salt. The continuity equation for air in the pores is unimportant in the presence of liquid moisture and can be omitted, assuming that the gas pressure in the pores $P_g$ is constant and equal to the external pressure $P_{g,ex}$. The heat transfer equation for a sphere (ignoring the gas phase in the pores) remains in the system for modeling non-isothermal drying.

As a result of the simplifications and assumptions made from the system of governing equations presented in \cite{RPK1,RPK2}, the system (1)-(4) below is obtained.
This system describes the change in the fraction of the sphere's volume occupied by the liquid, $\epsilon_l(r,t)$, the mass fraction of the metal salt in the liquid, $\gamma_m(r,t)$ ($\gamma_w=1-\gamma_m$), the fraction of the sphere's volume occupied by the precipitate, $\epsilon_m(r,t)$, and the temperature $T(r,t)$ along the radial coordinate $r$ at time $t$. It is assumed that all phases are in thermal equilibrium, and the temperature of both the solid and liquid phases is the same.

\begin{eqnarray}
\label{we}
&&\frac{\partial}{\partial t}(\epsilon_l\rho_l\gamma_w) + \nabla\cdot(\rho_l\gamma_w\bm v_l)  = \nabla\cdot(\epsilon_l\rho_lD_{wm}\nabla\gamma_w), \\
\label{se}
&&\frac{\partial}{\partial t}(\epsilon_l\rho_l\gamma_m) + \nabla\cdot(\rho_l\gamma_m\bm v_l) = \nabla\cdot(\epsilon_l\rho_lD_{wm}\nabla\gamma_m) - \epsilon_lr_{pr}, \\
\label{me}
&&\frac{\partial}{\partial t}(\epsilon_m\rho_m) = \epsilon_lr_{pr}, \\ 
\label{te}
&&\frac{\partial}{\partial t}(\epsilon_s\rho_sc_s + \epsilon_l\rho_lc_l + \epsilon_m\rho_mc_m)T + \nabla\cdot(\rho_lc_lT\bm v_l) = \nabla\cdot(\lambda\nabla T).
\end{eqnarray}

Equations (1) and (2) are the continuity equations for the aqueous and salt fractions of the liquid, respectively, taking into account binary diffusion in the liquid phase. Equation (3) describes the accumulation of immobile precipitated salt in the pores. The precipitation rate, $r_{pr}$, is defined as 
\begin{equation}
r_{pr} = k_{pr}\max(0,\gamma_m-\gamma_m^*),
\end{equation}
where $k_{pr}$ is the rate constant, $\gamma_m^*$ is the mass fraction of salt in a saturated solution. Equation (4) is the heat transfer equation in a three-component medium: solid skeleton (s), liquid (l), and precipitated salt (m), taking into account the flow of the liquid and the thermal conductivity of the medium.

\subsection{Parameters}

{\it Constant parameters} of the model: 
\begin{itemize}
\item[$\epsilon_s$] is the fraction of the material volume occupied by the solid skeleton, $\epsilon_s = 1 - \epsilon$. In this paper, $\epsilon$ = 0.6, so $\epsilon_s$ = 0.4.
\item[$\rho$] [kg/m$^3$] is the density of substance. Solid skeleto, $\rho_s$ = 3000; solt, $\rho_m$ = 1890; liquid, $\rho_l$ = 1050. 
\item[$c$] [J/(kg$\cdot$K)] is the specific heat capacity (at constant pressure). $c_s$ = 960; $c_m$ = 2610; $c_l$ = 4190.
\item[$D_{wm}$] [m$^2$/s] is the coefficient of binary diffusion of dissolved salt in water. $D_{wm}$ = $2\cdot 10^{-9}$.
\item[$\lambda$] [W/(m$\cdot$K)] is the thermal coductivity coefficient. $\lambda_s$ = 0.6; $\lambda_l$ = 0.8. The thermal conductivity of a wet sphere is estimated as
\begin{equation}
\lambda = \epsilon_s\lambda_s + \epsilon_l\lambda_l.
\end{equation}
\item[$k_{pr}$] [kg/(s$\cdot$m$^3$)] is the precipitation rate constant. The values of $k_{pr}$ are given in the figure captions. 
\end{itemize}
It is assumed that the liquid density, heat capacity and thermal conductivity are independent on solt concentration. If necessary, these dependencies can be easily taken into account. The main goal of this paper is to demonstrate the model, not to calculate specific examples. Therefore, the parameter values are rather arbitrary.

{\it Variable parameters} of the model are the surface tension, the dynamic viscosity and the soluble threshold of salt in water. Formulas for assessing the physical properties of water vapor and water depending on temperature [°C] are taken from works \cite{RPK1, RPK2}. \\
Pressure of saturated water vaper $P_{vp}^*$ [Pa]:
\begin{equation}
P_{vp}^*(T) = 133.32\exp\left(18.584-\frac{3984.2}{233.426+T}\right). 
\end{equation}
Surface tension of water $\sigma_w$ [N/m]:
\begin{equation}
\sigma_w(T) = -1.3\cdot 10^{-7}T^2 - 1.58\cdot 10^{-4}T + 0.07606.
\end{equation}
Dynamic viscosity of water $\mu_w$ [Pa$\cdot$s]:
\begin{equation}
\mu_w(T) = -1.27\cdot 10^{-9}T^3 + 3.42\cdot 10^{-7}T^2 - 3.43\cdot 10^{-5}T + 1.56\cdot 10^{-3}.
\end{equation}

The mass fraction $\gamma_m^*$ of salt in a saturated solution determines the conditions for precipitation. As an example of the dependence of this parameter on temperature, data on the solubility of ferrous sulfate in water, taken from Wikipedia (https://en.wikipedia.org/wiki/Iron(II)\_sulfate), are used in this work. The dependence of solubility $S_m^*$ [g per 100ml of water] on temperature in the temperature range of 0-60°C is approximated by the function
\begin{equation}
S_m^*(T) = 0.665T + 14.128,
\end{equation}
so (assuming the water density is equal 1000 kg/m$^3$) the limited mass fraction is defined as
\begin{equation}
\gamma_m^*(T) = \frac{S_m^*(T)}{100+S_m^*(T)}.
\end{equation}

The surface tension and viscosity of a liquid should depend on the salt concentration. However, information on these relationships is very scarce. Therefore, for the sake of certainty, it is assumed that in a saturated solution
\begin{equation}
\sigma_L^* = k_\sigma \sigma_w, \; \mu_l^* = k_\mu \mu_w,
\end{equation}
and a linear dependence in between
\begin{equation}
\sigma_l = \sigma_w + \frac{\gamma_m}{\gamma_m^*}(\sigma_l^*-\sigma_w), \;
\mu_l = \mu_w + \frac{\gamma_m}{\gamma_m^*}(\mu_l^*-\mu_w).
\end{equation}
In further calculations $k_\mu$ = 10 and $k_\sigma$ =1.

\subsection{Velocity of liquid}

The velocity $\bm v_l$ of the liquid is determined by Darcy's law
\begin{equation}
\bm v_l = - \frac{k_l}{\mu_l}\nabla P_l,
\end{equation}
where $k_l$ [m$^2$] is the permeability of porous material for liquid and $\mu_l$ is the dynamic viscosity of liquid. The liquid pressure $P_l$ is definrd as 
\begin{equation}
P_l = P_g - P_c,
\end{equation}
where $P_g$ is the gas pressure in the pores, and $P_c$ is the capillary pressure of the liquid. Since the gas phase is not taken into account in the model, the gas pressure $P_g$ is assumed to be constant and equal to the gas pressure $P_{g,ex}$ in the drying environment.Therefore, in fact,\begin{equation}
\label{vl2}
\bm v_l =  \frac{k_l}{\mu_l}\nabla P_c.
\end{equation}

The permeability of a porous material is determined by the structure of its pores. In this case, the density function $dV_p/dr_p$ of the pore volume $V_p$ distribution by pore size (radius) $r_p$ plays an important role. Like many papers on this topic, it is assumed that the distribution density has the form of a truncated Gaussian density
\begin{equation}
\label{dVp}
\frac{dV_p}{dr_p} = \frac{C}{\sigma_p\sqrt{2\pi}} \exp\left[-\frac{1}{2}\left(\frac{r_p-\overline{r_p}}{\sigma_p}\right)^2\right],\; r_p \in (r_{p,1},r_{p,2}),
\end{equation}
where $r_{p,1}$ and $r_{p,2}$ are the minimum and maximum pore radii, $\overline{r_p}$ is the average radius, and $\sigma_p$ is the standard deviation (in a Gaussian distribution). $C$ is a normalization constant chosen so that the integral of the density over the interval $(r_{p,1},r_{p,2})$ (the total pore volume) is equal to the porosity $\epsilon$.

To calculate the velocity of a liquid, the model of a porous medium as a bundle of capillaries is used with the volume distribution by radius (\ref{dVp}). This model assumes that when the medium is soaked, the liquid fills the capillaries in order of increasing radius, and when the liquid dries, it leaves them in the reverse order. Thus, the volume occupied by the liquid can be represented as follows:
\begin{equation}
\label{epl}
\epsilon_l = \int_{r_{p,1}}^{r_{p,f}}{\frac{dV_p}{dr_p}\,dr_p},
\end{equation}
where $r_{p,f}$ is the largest radius of liquid-filled capillaries. For known $\epsilon_l$, formula (\ref{epl}) is considered as an equation for unknown $r_{p,f}$. By solving this equation, one can find $r_{p,f}(\epsilon_l)$ and calculate the pressure in a capillary with a moving liquid using Young-Laplace formula.
\begin{equation}
\label{pce}
P_c(\epsilon_l) = \frac{2\sigma_l\cos(\theta)}{r_{p,f}(\epsilon_l)},
\end{equation}
for zero contact angle $\theta$ = 0. Using the capillary bundle model the permeability of a material to liquid can be estimated as
\begin{equation}
\label{kl}
k_l = \frac{1}{8}\int_{r_{p,1}}^{r_{p,f}}{r_p^2\frac{dV_p}{dr_p}\,dr_p}.
\end{equation}

Formulas (\ref{vl2}-\ref{kl}) provide the liquid velocity distribution $\bm v_l(r,t)$  for any instant of time $t$.

\subsection{Boundary and initial conditions}

The boundary conditions for equations (\ref{we}), (\ref{se}), (\ref{te}) are specified at the center of the sphere at $r=0$ and at the surface at $r=R$. At $r=0$, zero fluxes of the system variables are specified. At $r=R$, the flow of water is specified in the form of water vapor evaporating from the surface of the sphere:
\begin{equation}
\label{jv}
j_v = \epsilon_l(r,t)\beta P_{g,ex}\frac{M_v}{R_gT_{ex}} \ln\left(\frac{P_{g,ex}-P_{v,ex}}{P_{g,ex}-P_v^*(r,t)}\right)\bigg|_{r=R},
\end{equation}
where $\beta$ [m/s] is the external  mass transfer coefficient ($\beta$ = 0.015 \cite{RPK1}), $M_v$ (= 18 g/mol) is the molar weight of water vape, $R_g$ is the ideal gas constant. The flow of $\gamma_m$ at $r=R$ is zero. 

The heat flow at $r=R$ consists of two parts: the heat exchange term with the coefficient $\alpha$ [W/(m$^2\cdot$K)] ($\alpha$ = 14.25) and the heat transferred by vapor flow $j_v$
\begin{equation}
\label{jh}
j_h = \alpha(T_{g,b}(t)-T)+\Delta h_vj_v,
\end{equation}
where $\Delta h_v$ = -2.5$\cdot 10^6$ J/kg is the enthalpy of the water vapor. $T_{g,b}$ is the temperature of the external gas at the surface of the sphere. This temperature is introduced to avoid computational difficulties when $t$ = 0.
\begin{equation}
T_{g,b}(t) = \min(T_0+k_bt,T_{g,ex}), \; \mathrm{at}\, T_0\leq T_{g,ex},
\end{equation}
where $k_b$ is the rate of temperature change. In this paper, $k_b$ = 5 °C/s.

Initial conditions: the pores are completely filled with a liquid with a uniform distribution of the mass fraction of salt and with a uniform distribution of temperature:
\begin{equation}
\label{ic}
\epsilon_l(r,0) = \epsilon, \; \gamma_m(r,0) = \gamma_{m,0}, \; T(r,0) =T_0.
\end{equation}
At $t = 0$ there is no precipitant in the pores:
\begin{equation}
\epsilon_m(r,0) = 0.
\end{equation}

Note that the state 
\begin{equation}
\label{ss}
\epsilon_l(r) = 0,\; \gamma_m(r) = \gamma^*_m(T_{ex}), \; T(r) = T_{ex},
\end{equation}
is a stable stationary state of system (\ref{we})-(\ref{te}) with the specified boundary conditions (\ref{jv}), (\ref{jh}). Due to the uniqueness theorem, the solution with the initial condition (\ref{ic}) will asymptotically approach the stationary state (\ref{ss}) at $t \to \infty$. (Strictly speaking, the liquid in the sphere will never dry completely.) Therefore, solving the problem is stopped when the following condition is satisfied
\begin{equation}
\label{ed}
\max_r\epsilon_l(r,t) < \delta,
\end{equation}
for sufficiently small $\delta$.

\subsection{Numerical method}

Since most of the terms in equations (\ref{we})-(\ref{te}) have a divergent form, it is natural to use the control volume method for the numerical solution of the system.

To implement this method, the sphere is divided into $N$ spherical layers by spherical surfaces of radius $r_k=(k/N)^{1/3}R$, $k = 1,2,.. ,N$ (the results of calculations with $N$ = 100 are presented below). The volume of each layer equals to $V/N$, where $V$ is the volume of the spherical pellet. The equations (\ref{we})-(\ref{te})  are integrated over each spherical layer, with the divergent terms being integrated exactly and the mean value theorem being used to estimate the integrals of the non-divergent terms. As a result of integration, a 4$N$ ODE system is obtained, which, after taking into account the boundary conditions, is solved using the MATLAB function `ode15s'. All results presented below were obtained with a relative tolerance RelTol = 10$^{-6}$ and an absolute tolerance AbsTol = 10$^{-9}$.

\section{Results}

\subsection{General view of the model solution}

\begin{figure}[h]
\centering
\includegraphics{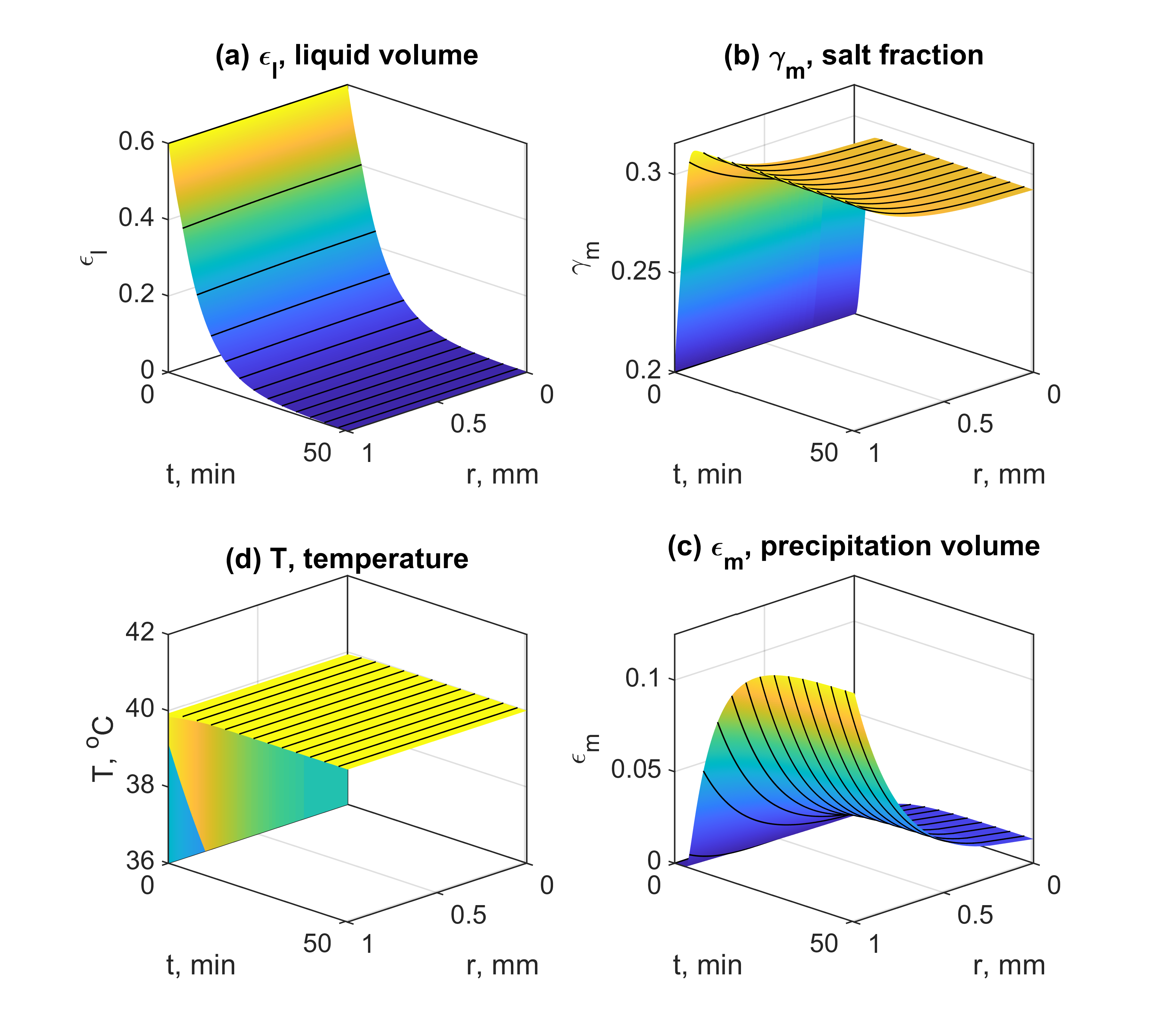}
\label{fg1}
\caption{Drying at $T_{ex}$ = 40°C. Initial salt mass fraction in liquid $\gamma_{m,0}$ = 0.2. Rate constant of precipitation $k_{pr}$ = 50 kg/(s$\cdot$m$^3$). Drying time -- 52 min. Lines on the surfaces were drawn at 4-minute intervals.}
\end{figure}

A general overview of a typical solution to system (1)–(4) is presented in Figure \ref{fg1}. This figure shows the space-time graphs of four system variables for solving the initial-boundary value problem under the following conditions. The sphere was impregnated with a liquid at a temperature of $T_0$ = 20°C. The mass fraction of salt in the liquid is $\gamma_{m,0}$ = 0.2. Drying occurs in air at atmospheric pressure ($P_{g,ex}$ = 101325 Pa) and a temperature of $T_{ex}$ = 40°C. The water vapor pressure in the drying air is constant and equal to $P_{v,ex}$ = 2333 Pa (which corresponds to 10\% air humidity at a temperature of 20°C). This pressure was the same in all the examples below. The numerical solving of the problem was terminated when condition (\ref{ed}) with $\delta$ = 0.001 was satisfied. Let us consider the behavior of each variable during drying.

Subplot (a). A graph of $\epsilon_l(r,t)$ is shown. During the initial drying stage, over the course of approximately 20 minutes, the fraction of the sphere's volume occupied by liquid rapidly decreases from an initial value of 0.6 to approximately 0.05 due to water evaporation and salt precipitation. This is followed by a slow drying stage, which lasts approximately 30 minutes, until condition (1) is satisfied. At each $t$, $\epsilon_l$ decreases monotonically with $r$. At the surface of the sphere, $\epsilon_l$ is slightly smaller than at the center.

\clearpage

Subplot (b). A  graph of the mass fraction of salt $\gamma_m(r,t)$ in the liquid is shown. During the first 5 minutes of drying, the mass fraction of salt increases from an initial value of 0.2 to a saturation value of 0.29 at a given temperature $T_{ex}$ = 40°C and then remains virtually unchanged until the end of drying. Moreover, at the surface of the sphere $\gamma_m$ is slightly larger than in the center.

Subplot (c).A graph of the fraction of the sphere's volume $\epsilon_m(r,t)$ occupied by salt precipitation is shown. It is clear that significantly more precipitation falls in the sphere's surface layers than in its center. The reasons for this phenomenon are discussed in Section 3.2.. 

Subplot (d). The temperature graph $T(r,t)$ is shown. The temperature remains virtually constant throughout the drying process, with the exception of a jump from $T_0$ to $T_{ex}$ at the very beginning of the process.
%\clearpage

\begin{figure}[h]
\centering
\includegraphics{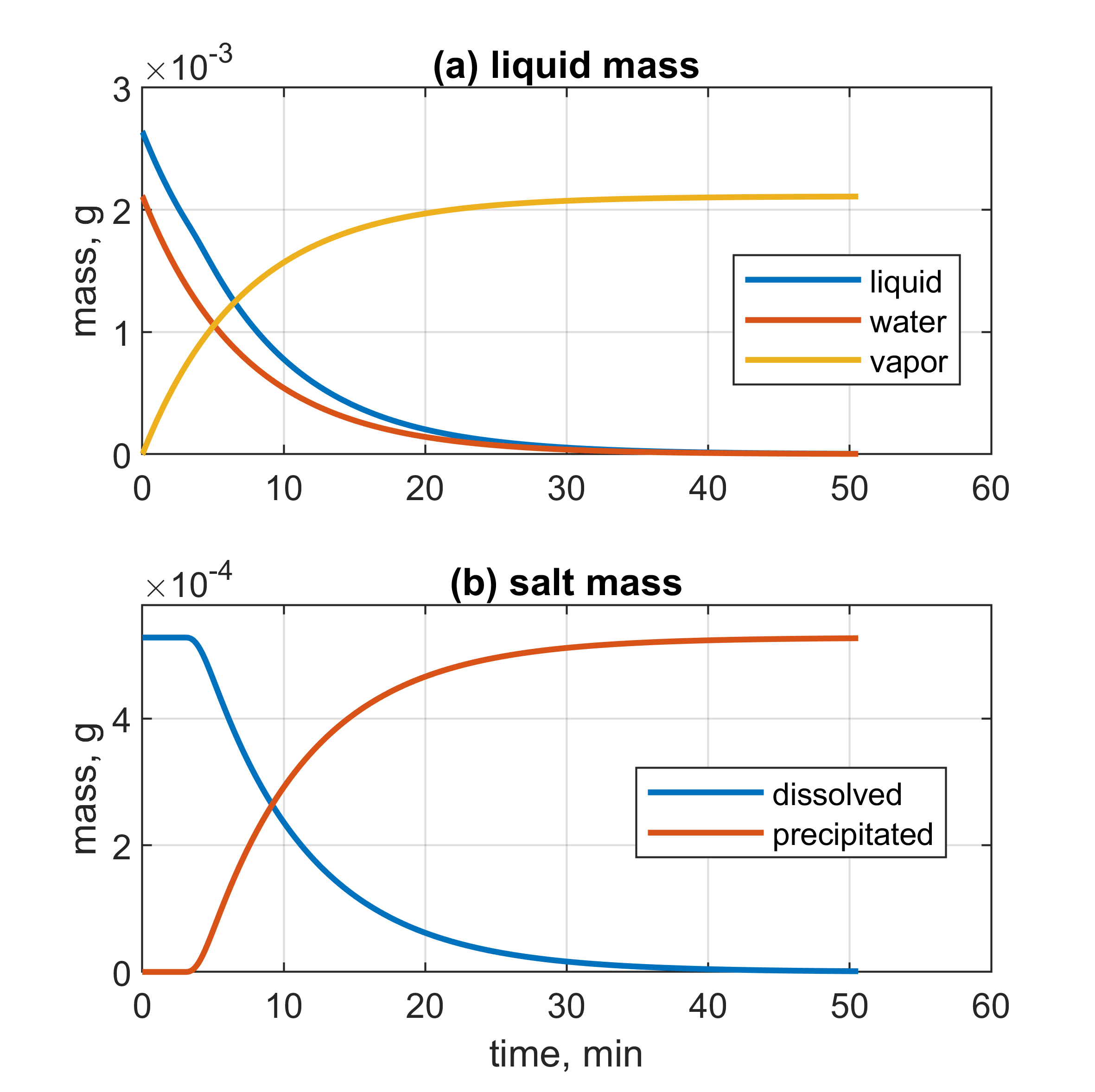}
\label{fg2}
\caption{Change in the total mass of substances during drying.}
\end{figure}

The total change in the mass of the components involved in the process during drying is shown in Figure \ref{fg2}.

\subsection{Details of precipitation}

\begin{figure}[h]
\centering
\includegraphics{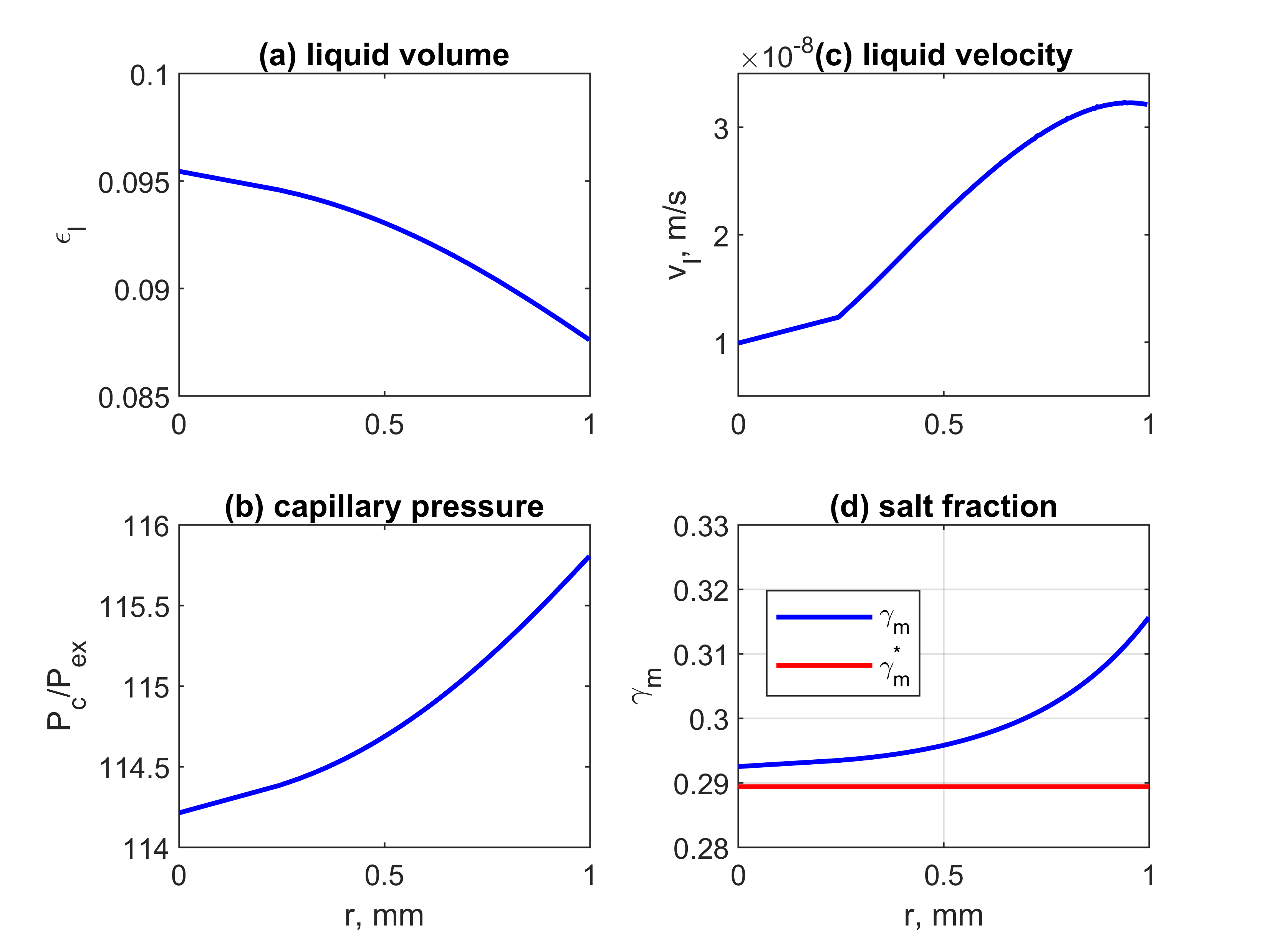}
\label{fg3}
\caption{Details of the solution of system (1)-(4) at time $t_*$ = 15 min.}
\end{figure}

Figure \ref{fg3} shows some details of the solution of system (1)-(4), explaining the features of the process of salt precipitation. Shown here are the graphs relating to the time $t_*$ = 15 min from the problem  solution, the general form of which is shown in Fig. \ref{fg1}.

From the graph of $\epsilon_l(t_*,r)$, shown in subplot (a), it follows that the volume occupied by the liquid decreases from the center of the sphere to its periphery. Therefore, the radius $r_{p,f}$ (eq. (\ref{epl})) of the largest capillary filled with liquid also decreases and, accordingly, the capillary pressure $P_c$ (eq. (\ref{pce})) increases. The graph of capillary pressure $P_c(t_*,r)$ is shown in subplot (b) in relative dimensionless units $P_c/P_{ex}$, where $P_{ex}$ is the atmospheric pressure.

An increase in the function $P_c(r)$ leads to a positive liquid velocity $v_l(r)$, the graph of which is shown in subplot (c). The llquid flows from the center to the surface of the sphere, where water evaporates from the liquid. The water evaporates, but the salt remains in liquid, increasing the salt concentration at the surface of the sphere (subplot (d)). Therefore, the rate of precipitation is greater at the surface of the sphere than at the center.

\subsection{ Precipitation distribution over radius}

A practically important output of the model is the ability to simulate the influence of various factors on the distribution of precipitation along the radius of the sphere after the drying stage.

\begin{figure}[h]
\centering
\includegraphics{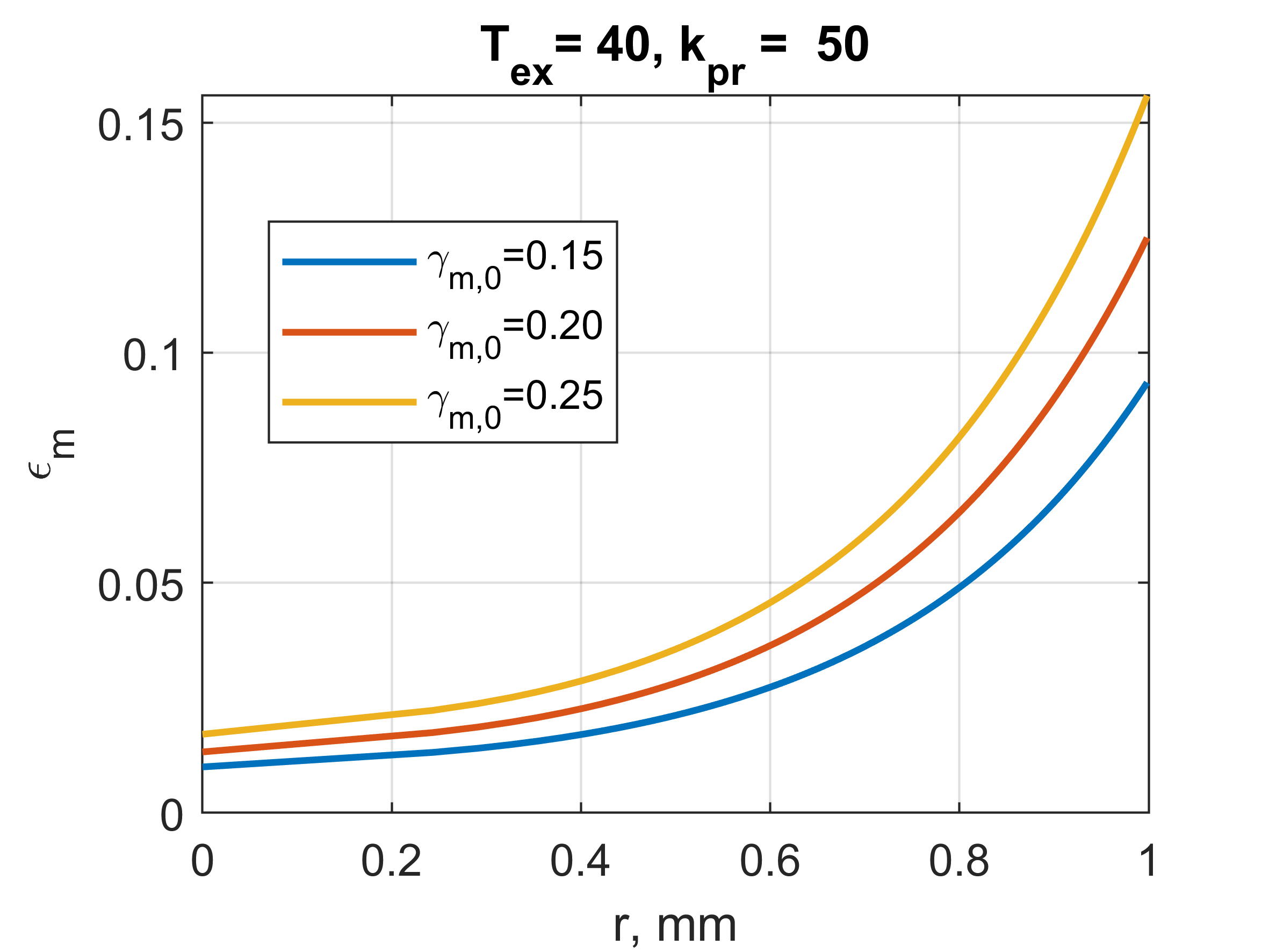}
\label{fg4}
\caption{Distribution of precipitation by radius at the end of drying depending on the initial mass fraction of salt.}
\end{figure}

Figure \ref{fg4} shows the graphs of the distribution of the salt volume by the radius of the sphere after drying at a temperature of 40°C. The graphs are constructed based on the results of solving the system (1)-(4) with different initial concentrations of salt in the aqua. With increasing mass of salt in the aqua, the lines in the figure rise, maintaining their shape.

The following Figure \ref{fg5} demonstrates the influence of the drying temperature $T_{ex}$ and the value of the salt precipitation rate constant $k_{pr}$ on the final distribution of the precipitated salt.

Figure \ref{fg5}a shows that the final salt distribution is virtually independent of drying temperature. Despite the significant differences in drying time: 165 minutes at $T_{ex}$ = 20°C, 52 minutes at 40°C, and 20 minutes at 60°C, the salt distribution curves are almost identical. Note, however, that at all three temperatures $T_{ex}$ the value of the constant $k_{pr}$ was the same and equal to 50.

\begin{figure}[h]
\centering
\includegraphics{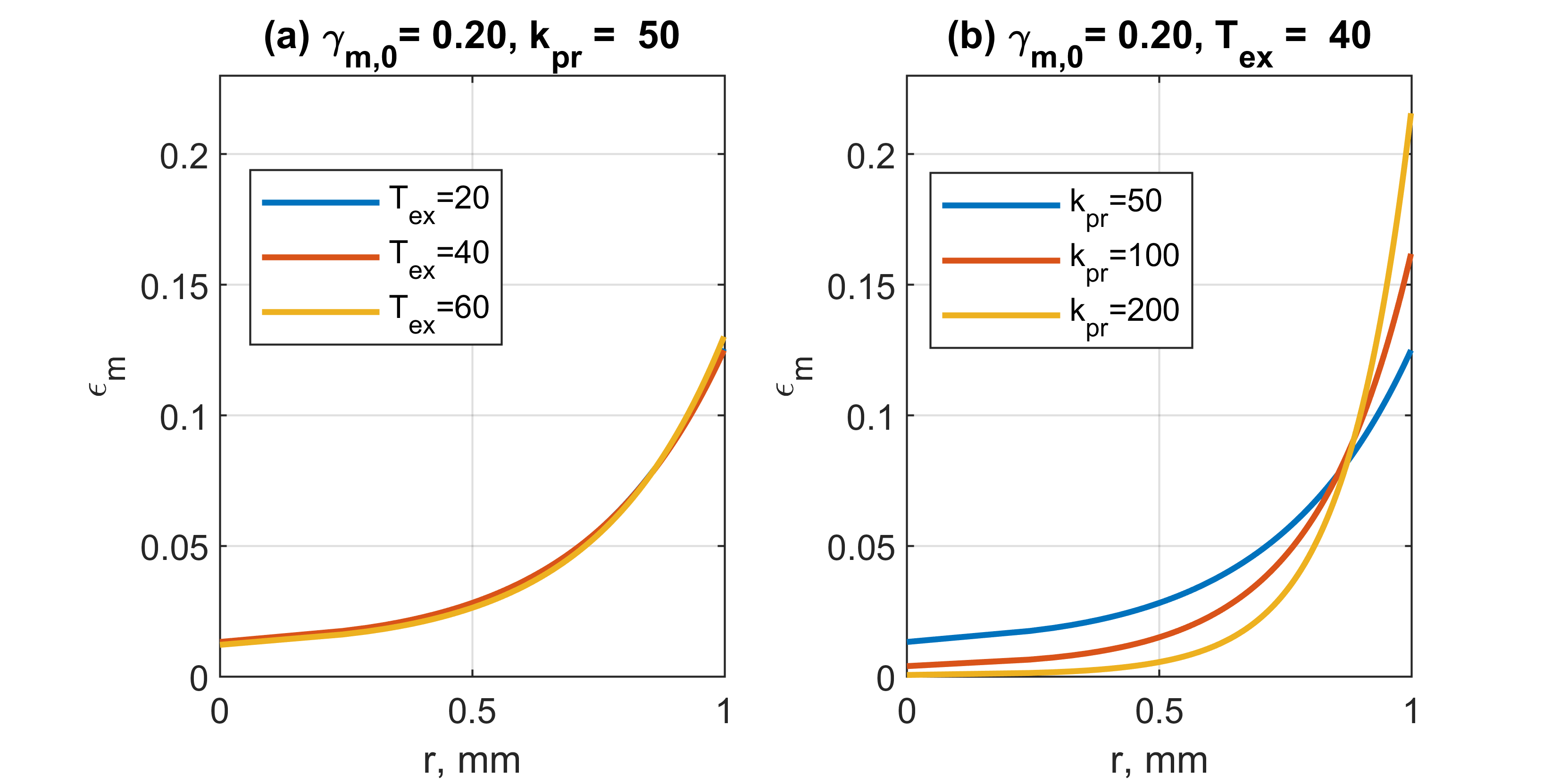}
\label{fg5}
\caption{Distribution of precipitation by radius at the end of drying depending on (a) the drying temperature $t_{ex}$, (b) the value of constant $k_{pr}$.}
\end{figure}

Figure \ref{fg5}b shows the significant influence of the parameter $k_{pr}$ on the distribution of precipitated salt after the liquid dries. As this constant increases, an increasing volume of salt precipitates in the near-surface region of the sphere and an increasingly smaller volume in its central region. If one assumes that the parameter $k_{pr}$ increases with increasing temperature, then a similar dependence of the distribution on the drying temperature should be observed. As the temperature increases, more and more salt will precipitate near the surface of the sphere.

\subsection{Distribution of precipitation by pore size}

\begin{figure}[h]
\centering
\includegraphics{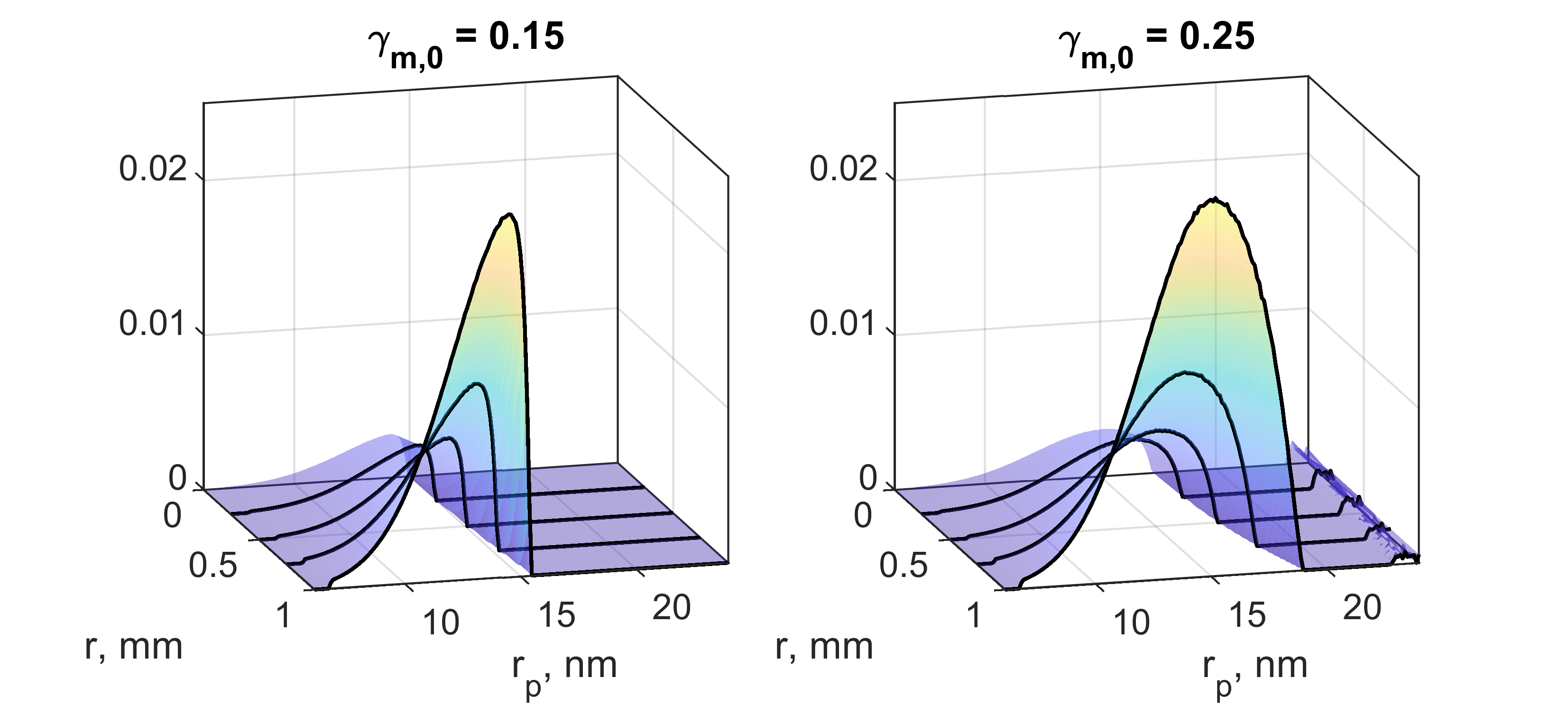}
\label{fg6}
\caption{The density of 2D distribution by the radius of the sphere and the pore size of the fraction of the sphere volume occupied by the precipitation after drying at a temperature of $T_{ex}$ = 40 and $k_{pr}$ = 50. The density is normalized to the total volume of salt,}
\end{figure}

The capillary bundle model, which is used in the Whitaker model to calculate capillary pressure and liquid velocity, assumes that the liquid in the pores dries in order of decreasing pore size. If one adds another assumption to this --  namely, that the precipitation occurs in the pores that are currently drying -- then the volume fraction occupied by precipitate distribution by radius and pore size can be easily calculated. An example of such a distribution is shown in Figure \ref{fg6}.

Salt deposition in pores causes changes in the pore volume distribution by size. These changes are particularly noticeable in the near-surface region of the sphere, where (in the examples discussed here) more salt is deposited. Note that these changes do not affect the results presented above, since the integration in formulas (\ref{epl}), (\ref{kl}) is carried out over capillaries filled with liquid, and in them (according to the assumption made) there is no precipitated salt yet, therefore the distribution $dV_p/dr_p$ at these sizes has not yet changed. However, these changes should be taken into account when simulating re-impregnation of the same pellet with a different precursor solution.

\begin{figure}[h]
\centering
\includegraphics{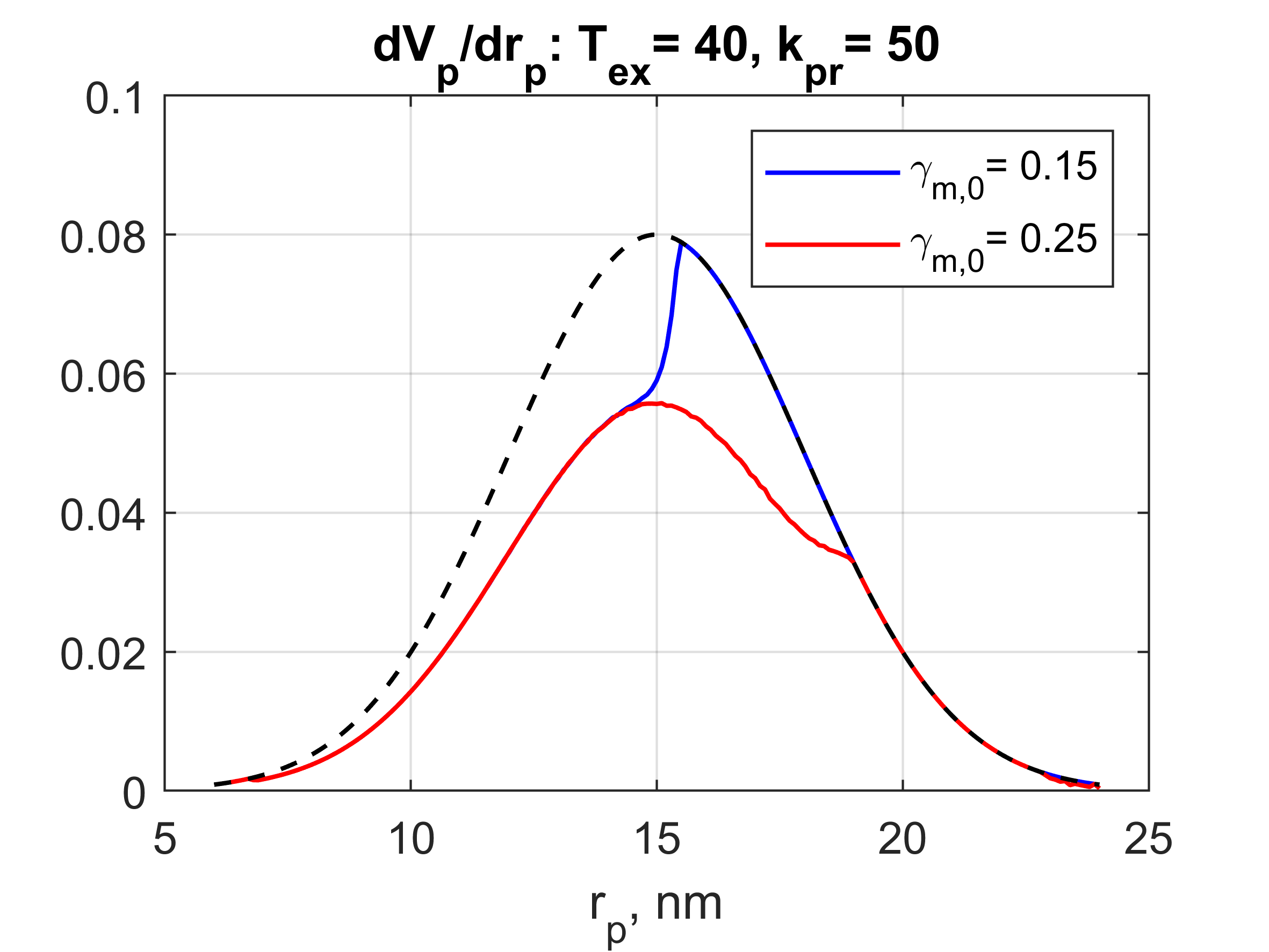}
\label{fg7}
\caption{Estimation of the pore volume distribution over the pore size in the near-surface layer of the sphere after drying. In large pores, no precipitate forms because the salt concentration in the solution is less than critical. The dashed line shows the original distribution.,}
\end{figure}

To assess the change in distribution $dV_]/dr_p$ during drying, one can subtract the distribution of precipitation volume by pore size from the initial distribution. An example of the result of such an operation is shown in Figure \ref{fg7}. This is only an approximate estimate, since it does not take into account the possible change in pore size when salt falls into them. In addition, during subsequent steps to prepare the catalyst, the salt must turn into a catalytically active substance with a different volume.

\section{Conclusion}

The above-presented mathematical model of drying an aqueous solution of a catalyst precursor in a porous sphere with precipitation of the dissolved substance in the pores is essentially a modified version of the model described in \cite{RPK1,RPK2}. The main modifications are as follows.

Firstly, in accordance with the goal of this work: to model the distribution over the volume of a sphere of precipitation falling from a liquid during drying, the presented model describes the drying process only in the presence of liquid in the pores, i.e. at $\epsilon_l(r,t) > 0$. Due to the boundary condition (\ref{jv}), any solution of system (1)-(4) with the initial state $\epsilon_l(r,0)>0$ asymptotically tends to the state in which $\epsilon_l(r,t)=0$ while remaining positive, $\epsilon_l(r,t)\to +0$ at $t\to\infty$. Thus, in this model, liquid is present in the pores throughout the entire drying period. Dynamic equilibrium between liquid and water vapor is assumed, i.e., the water vapor density in the pores $\rho_v$ is always equal to the saturated vapor density $\rho_v^*$.

Secondly, since the value of $\rho_v^*$ under the conditions under consideration is more than three orders of magnitude less than the density of the liquid, the model does not take into account the transfer of mass and heat in the gas phase in the pores. It is assumed that the gas pressure in the pores is always equal to the pressure of the external gas $P_{gmex}$.

The Whitaker drying model uses a representation of the porous medium as a bundle of capillaries of varying sizes. This representation establishes the order of pore drying; the liquid in the pores dries in descending order of pore size. Assuming that the dissolved salt precipitates in the pores that are currently drying, the precipitate distribution by pore size can be estimated. 

This paper demonstrates the influence of certain model parameters on the final distribution of precipitated salt across the volume of a sphere. The resulting distributions increase monotonically from the center to the surface of the sphere. A detailed parametric analysis of the model could reveal other distribution shapes It can be assumed that the shape of the final distribution is primarily determined by the hydrodynamics of the process, or more precisely, the liquid velocity distribution $v_l$. In this model, two factors influence liquid velocity: the surface tension of the liquid $\sigma_l$ and the pore volume distribution by size $dV_p/dr_p$, which were not varied in this study.

\section*{Acknowledgments}
 The author thanks Professor P.A. Chernavskii for helpful discussions.
  \\
   \\
 %\section*{Funding}
 This research did not receive any specific grant from funding agencies in the public, commercial, or not-for-profit sectors.

\end{document}